\documentclass[conference]{IEEEtran}
\IEEEoverridecommandlockouts
\usepackage{cite}
\usepackage{amsmath,amssymb,amsfonts}
\usepackage{amssymb}
\usepackage{algorithmic}
\usepackage{graphicx}
\usepackage{textcomp}
\usepackage{xcolor}
\usepackage{cite}
\usepackage{amsmath,amssymb,amsfonts}
\usepackage{algorithmic}
\usepackage{graphicx}
\usepackage{textcomp}
\usepackage{times}
\usepackage{xcolor}
\usepackage{float}
\usepackage{algorithm}
\usepackage{multirow}
\newcommand\independent{\protect\mathpalette{\protect\independenT}{\perp}}
\def\independenT#1#2{\mathrel{\rlap{$#1#2$}\mkern2mu{#1#2}}}
\usepackage{mathtools}
\usepackage{mathtools, nccmath}
\usepackage[font=small,skip=0pt]{caption}
\usepackage{fancyhdr}
\usepackage{hyperref}
\usepackage{makecell}

\def\BibTeX{{\rm B\kern-.05em{\sc i\kern-.025em b}\kern-.08em
    T\kern-.1667em\lower.7ex\hbox{E}\kern-.125emX}}

\begin{document}

\title{Optimization Results for 5G Slice-in-Slice Scheduling \\
\thanks{This research work has been done in the field of 5G Scheduling for network slicing by Sharvari Ravindran, Saptarshi Chaudhuri, Jyotsna Bapat, and Debabrata Das, IIIT Bangalore}
}

\author{\IEEEauthorblockN{Sharvari Ravindran, Saptarshi Chaudhuri, Jyotsna Bapat, and Debabrata Das}
\IEEEauthorblockA{\textit{Networking and Communication Research Lab} \\
\textit{International Institute of Information Technology, Bangalore, India}\\
Sharvari.R@iiitb.org, saptarshi.chauduri@iiitb.org, jbapat@iiitb.ac.in, ddas@iiitb.ac.in
}
}


\maketitle

\begin{abstract}
Open Radio Access Network (ORAN) Slicing for 5G and Beyond is an emerging architecture and feature that will facilitate challenging RAN Service Level Agreement (SLA) assurance targets. This could pave the way for operators to realize the benefits of network slicing efficiently. In this paper, we provide novel and detailed optimization results to achieve Slice-in-Slice Scheduling for 5G User Services in ORAN slicing.
\newline
\end{abstract}

\section{Result 1}
\textbf{Result 1:} \textit{The network slice capacity is infinitely small as the relative system resource allocation for user services across slice-in-slice categories $\rightarrow$ zero under certain necessary conditions.} \\
\textbf{Design model}: \textit{Throughput and system resources for increasing user services is learnt conditioned on previously occured and existing services within the network slice.} \\
\hspace{0.1cm}
Let $i$ denote the user services within the network slice, $m$ denote the slice-in-slice categories within the slice. Each slice-in-slice category within a network slice hosts user services. We aim to maximize the capacity of each slice-in-slice category, i.e., maximize the number of user services as,
\begin{equation}
C_{m} = max \; A_{m} \; \forall \; m = 1,2,...,S
\end{equation}
\hspace{0.1cm}
where $A_{m}$ are the maximum number of user services $\in m$. The capacity disregards effects of diversity from combining several resource pools. In this section, we aim to prove that the combined capacity tends to $0$ as the relative system resource per user service tends to $0$. Consider the total available system resources as $r_{max}$. Let the system resource allocation for each user service be defined by a random variable $R_{i}$. Let the expected valuser $E[R_{i}] = r$. The relative resource allocation is denoted as $A_{i}$ which would be defined as the ratio of system resources allocated to the total available system resources, i.e., $\frac{R_{i}}{r_{max}}$. Let $E[A_{i}] = a$. Let the number of user services be $U$ where $E[U] = s$. We now analyze the system metric performances. Let the total resource allocation ($T$) and relative resource allocation ($R$) be defined as,
\begin{equation}
T = \sum_{i=1}^{U} R_{i} \leq r_{max}, R = \sum_{i=1}^{U} \frac{R_{i}}{r_{max}} \leq 1
\end{equation}
\hspace{0.1cm}
We design a system such that the system resource allocation for user services is dependent on each other, i.e., dependency on previous user service(s) system resources (as per the design model). 
\hspace{0.1cm}
We assume that the dependencies are bounded using the mean, variance and covariance function given below,
\begin{equation}
\mathcal{B} \rightarrow
\begin{cases}
E[A_{i}|U] \leq a' \\
E[A_{i}|U] \leq \sigma_{a}^{2} \\
Cov[A_{i}, A_{j}|U] \leq c_{a'} \; i \neq j
\end{cases}
\end{equation}
\hspace{0.1cm}
Additionally, we assume that the SLA degrades when $T$ exceeds $r_{max}$, i.e., the SLA bound can be defined as,
\begin{equation}
L = P[T \leq r_{max}] = P[\frac{T}{r_{max}} \leq 1] = P[E \leq 1]
\end{equation}
\hspace{0.1cm}
We express the network slice capacity as sum of the individual capacities of each slice-in-slice category,
\begin{equation}
\mathcal{C} = \sum_{m = 1}^{S} C_{m} =  \sum_{m = 1}^{S} max \; A_{m} 
\end{equation}
\hspace{0.1cm}
In general, each network slice can host a maximum number of user services ($s$) which would be bounded (depending on the available system resources) such that,
\begin{equation}
s \leq \frac{g . r_{max}}{r} \leq \frac{g}{a} 
\end{equation}
\hspace{0.1cm}
where $g > 1$ is a constant. The slice-in-slice category capacity is defined as the maximum expected number of user services,
\begin{equation}
\begin{split}
C_{max} & = max \; A_{m} = max \; A_{m}: P[T \leq r_{max}] = max \\ & \; A_{m}: P[E \leq 1] 
\end{split}
\end{equation}
\hspace{0.1cm}
At the maximum system capacity limit, it might be possible that the system has some amount of free available system resources, 
\begin{equation}
P[T \leq r_{max}] > 0 \rightarrow P[R \leq 1] > 0
\end{equation}
\hspace{0.1cm}
To show the validity of (8), we show that under certain conditions the variance of $E \rightarrow 0$ as the relative system resource allocation $\frac{R_{i}}{r_{max}} \rightarrow 0$. We solve for $V[R]$ which follows from the concept of Bienayme–Chebyshev inequality [1]. From the law of total variance, we know that $V(Y) = E[V[Y/X]] + V(E[Y/X])$,
\begin{equation}
V[R] = E[V[R|U]] + V[E[R|U]]
\end{equation}
\hspace{0.1cm}
Inserting $R = \sum_{i=1}^{U} \frac{R_{i}}{r_{max}} =  \sum_{i=1}^{U} A_{i}$ in (9), we notice the term $A_{i}$ which comprises of several user services $U$. Since we have considered dependency across the user services system resources, we re-express (9) as,
\begin{equation}
\begin{split}
V[R] & = E[\sum_{i=1}^{U} V[A_{i}|U] + \sum_{i=1}^{U} \sum_{j \neq i} Cov[A_{i}, A_{j}|U]] + \\ & V[E[\sum_{i=1}^{U} A_{i}|U]]
\end{split}
\end{equation}
\hspace{0.1cm}
Eqn. (10) is complex to solve. We use the bounds and limits discussed earlier in (3) to yield,
\begin{equation}
\begin{split}
V[R] & \leq E[\sum_{i=1}^{U} \sigma_{a}^{2} + \sum_{i=1}^{U} \sum_{j \neq i} c_{a'}] + V[\sum_{i=1}^{U} a']
\end{split}
\end{equation}
\begin{equation}
\begin{split}
V[R] = \sigma_{a}^{2}E[\sum_{i=1}^{U} 1] + c_{a'} E[\sum_{i=1}^{U} \sum_{j \neq i} 1] +  a'^{2} V[\sum_{i=1}^{U}]
\end{split}
\end{equation}
\hspace{0.1cm}
We know that $\sum_{i=1}^{U} 1 = 1 + 2 + 3 + ... + U = \frac{U(U+1)}{2}$. Further, since $j \neq i$, the summation limits would be from $\sum_{i=1}^{U-1} 1 = \frac{U(U-1)}{2}$. Inserting in (12),
\begin{equation}
\begin{split}
V[R] & = \sigma_{a}^{2}E[\frac{U(U+1)}{2}] + \frac{c_{a'}}{4} E[U^{2}(U+1)(U-1)] +  \\ & a'^{2}V[\frac{U(U+1)}{2}]
\end{split}
\end{equation}
\begin{equation}
\begin{split}
V[R] = \frac{\sigma_{a}^{2}}{2} E[U^{2} + U] + \frac{c_{a'}}{4} E[U^{4} - U^{2}] +  \frac{a'^{2}}{4} V[U^{2} + U]
\end{split}
\end{equation}
\hspace{0.1cm}
We further know that $E[A + B] = E[A] + E[B]$, $Var(A + B) = Var(A) + Var(B) + 2 Cov(A,B)$ which is inserted in (14) to yield,
\begin{equation}
\begin{split}
V[R] & = \frac{\sigma_{a}^{2}}{2} (E[U^{2}] + E[U]) + c_{a'} E[U^{4} - U^{2}] + \frac{a'^{2}}{4} [V[U^{2}] \\ & + V[U] + 2 Cov(U^{2}, U)]
\end{split}
\end{equation}
\hspace{0.1cm}
Solving, $V[U^{2}] = E[U^{4}] - E[U^{2}]^{2} = E[(\frac{g}{a})^{4}] - E[(\frac{g}{a})^{2}]^{2} = 0$ since $\frac{g}{a}$ is a constant. Similarly, $V[U] = 0$, as $U = \frac{g}{a}$ is a constant. Hence,
\begin{equation}
\begin{split}
V[R] & = \frac{\sigma_{a}^{2}}{2} ((\frac{g}{a})^{2} + \frac{g}{a}) + c_{a'} E[U^{4} - U^{2}] +  \frac{a'^{2}}{4} \\ & [2 Cov(E[U^{2}],E[U])]
\end{split}
\end{equation}
\hspace{0.1cm}
Thus, $V(R) \rightarrow 0$ as $a \rightarrow 0$, $a^{'} \rightarrow 0$ ($a < a'$) provided the below conditions are satisfied,
\begin{equation}
\begin{cases}
lim_{a \rightarrow 0} \frac{\sigma_{a}^{2}}{2} ((\frac{g}{a})^{2} + \frac{g}{a}) = 0 \\
lim_{a' \rightarrow 0} c_{a'} E[U^{4} - U^{2}] = 0 \\
lim_{a' \rightarrow 0} \frac{a'^{2}}{4} 2 Cov(E[U^{2}],E[U]) = 0 \\
\end{cases}
\end{equation}
\hspace{0.1cm}
Hence, the network slice capacity $\rightarrow 0$ for relative system resource allocation $a < a' \rightarrow 0$ provided conditions in (17) are satisfied. 

\section{Result 2}
\textbf{Result 2}: \textit{The estimation of user service throughput and system resources within a network slice} \textit{induces a maximal clique to form a perfect graph} \\
\textbf{Extended Result}: \textit{The formulated perfect graph uses $\prod_{n=1}^{U} (k-n+1)$ colors, where $k$ is maximum available colors}. \\
To prove the result, we decompose the problem into two parts. The first part deals with proving that the problem induces a maximal clique. The second part deals with proving that the resultant is a perfect graph. \par 
Let $1_{m}, 2_{m},...u_{m}$ denote the user services within $m^{th}$ slice-in-slice category. Let us assume there exists only 1 user service $1_{m}$ within the network slice. We can formulate our resultant graph as in Stage 1 of Fig. 1. Consider a second user service $2_{m}$ enters the network slice. Based on our design consideration, we need to estimate the throughput and system resources for $2_{m}$. Since $1_{m}$ already exists within the network slice, the throughput and system resources for $2_{m}$ is learnt conditioned on $1_{m}$. In probabilistic terms, we can express this as $P(2_{m}|1_{m}) = \frac{P(2_{m}, 1_{m})}{P(1_{m})}$ with dependency on $1_{m}$. The resultant graph for 2 user services is shown in Fig. 1. For increasing number of user services increases, the previously occured services ($\mathcal{V}$) are,
\begin{equation}
\begin{cases}
3_{m} \rightarrow \mathcal{V} = 1_{m}, 2_{m} \\
4_{m} \rightarrow \mathcal{V} = 1_{m}, 2_{m}, 3_{m} \\
5_{m} \rightarrow \mathcal{V} = 1_{m}, 2_{m}, 3_{m}, 4_{m} \\
... \\
u_{m} \rightarrow \mathcal{V} = 1_{m}, 2_{m}, 3_{m}, 4_{m}, 5_{m}, ...,u_{m}-1  \\
\end{cases}
\end{equation}
\hspace{0.1cm}
Based on (18), generalizing, we can state that $P(u_{m}| \mathcal{V}) = \frac{P(u_{m}, \mathcal{V})}{P(\mathcal{V})}$. The complete resultant graph for $u_{m}$ user services is shown in Fig. 1. Each user service is connected to the other services thus demonstrating the throughput and system resource dependency relation. To show that the graph induces a clique, we start with its definition. A clique is a subset of vertices of undirected graph where every two distinct vertices are adjacent [2]. As shown in Fig. 1, we observe the graph induces a maximal and maximum clique as, $(i)$ All vertices are connected to each other, $(ii)$ Graph cannot be extended by including an additional adjacent vertex, $(iii)$ $u_{m}$ are the maximum number of user services or vertices in the network. Thus, we denote the clique as $1-2-3-4-...-u$. We now move to the second part, i.e., proving the resultant graph is perfect. To show this, we have to show that the chromatic number of the graph equals the clique number. We have already shown the maximal clique previously. To evaluate the chromatic number, we need to color the vertices of the graph. Since each vertex is adjacent to each other using distinct edges, from the concept of graph coloring, each vertex has to be assigned a distinct color. 
 \begin{figure*}
\begin{center}
\includegraphics[scale = 0.23]{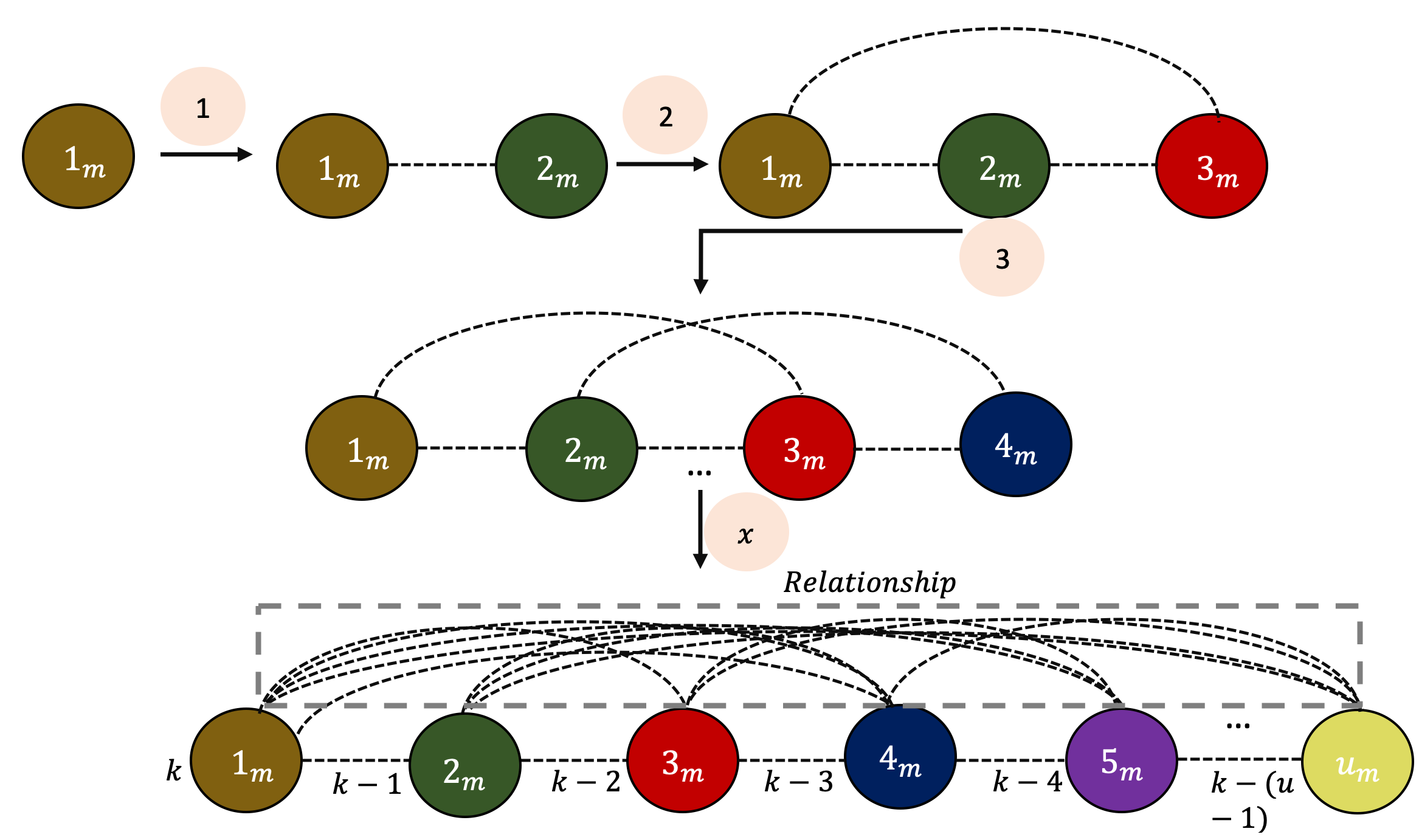}
\caption{Chromatic color polynomial for perfect graph}
\end{center}
\vspace{-5mm}
\end{figure*}
Thus, from the above analysis, we see that the chromatic number or the color is assigned distinctly to $\{1,2,3,4,...,u\}$ which equals the clique number of $1-2-3-4-...-u$. Once the vertices are colored, we now count the total number of colors used to achieve the optimization. Assume total of $k$ colors available. These colors denote the available resources (system resources) per network slice. Since each vertex is connected to each other using distinct edges, each vertex will be assigned a new color. Hence, if vertex $1_{m}$ can be colored in $k$ ways, vertex $2_{m}$ will be colored in $(k-1)$ ways,  vertex $3_{m}$ in $(k-2)$ ways and so on. Thus, an $n^{th}$ vertex would be colored in $(k-(n-1))$ ways. We can then represent the chromatic polynomial as,
\begin{equation}
\begin{split}
\mathcal{C} & = k(k-1)(k-2)(k-3)(k-4)(k-5)....(k-(s-1)) \\ & ...(k-(u-1)) = \prod_{n=1}^{u} (k-n+1) 
\end{split}
\end{equation}

\section{Result 3}
\textbf{Result 3}: \textit{An optimal perfect graph coloring strategy is bounded by} $\chi(G) = u_{m+k}$. \\
While performing coloring, it is vital to check whether the coloring scheme is optimal and bounded. In Result 2, we have shown that the total number of colors needed to color the graph equals the total number of vertcies in the graph, i.e., $u_{m}$. To prove if the result is bounded and optimal, we start with defining the degree of each vertex. In the graph of Fig. 1, we notice that each vertex has the same degree defined by the total number of vertices $- \; 1$, i.e., $u_{m} - 1$. We order the vertices based on their degree sequence as,
\begin{equation}
\{1_{m},2_{m},...,u_{m}\}
\end{equation}
\hspace{0.1cm}
Based on the ordering of vertices, we define the coloring strategy depending on the number of earlier neighboring vertices ($\mathcal{E}$) that have already been colored. 
\begin{equation}
\begin{cases}
1_{m}: \mathcal{E} = 0 \\
2_{m}: \mathcal{E} = 1 \\
3_{m}: \mathcal{E} = 2 \\
4_{m}: \mathcal{E} = 3 \\
... \\
u_{m}: \mathcal{E} = u_{m} - 1 \\
\end{cases}
\end{equation}
Based on $\mathcal{E}$, we can define in general the number of earlier neighbors associated with each vertex as a function of the vertex index and associated degree,
\begin{equation}
\begin{cases}
1_{m}: \mathcal{E} = min(u_{m} - 1,(1-1)) \\
2_{m}: \mathcal{E} = 1 = min(u_{m} - 1,(2-1)) \\
3_{m}: \mathcal{E} = 2 = min(u_{m} - 1,(3-1)) \\
4_{m}: \mathcal{E} = 3 = min(u_{m} - 1,(4-1))\\
... \\
u_{m}: \mathcal{E} = u_{m} - 1 = min(u_{m} - 1,u_{m} - 1)\\
\end{cases}
\end{equation}
\hspace{0.1cm}
Considering the final graph of Fig. 1, to color the last vertex $u_{m}$, we have,
\begin{equation}
\chi(G) = \chi(u_{m}) + \chi(1,2,...,(u_{m} - 1))
\end{equation}
\hspace{0.1cm}
One distinct color would be needed to color each vertex (as shown in Result 2),
\begin{equation}
\chi(G) = 1 + u_{m} - 1 = u_{m} 
\end{equation}
\hspace{0.1cm}
Thus (24) proves that the number of colors for $G$ is optimal and bounded by $u_{m} $.

\section{Result 4}
\textbf{Result 4}: \textit{The system resource allocation for each user service is learnt based on the total probability distribution functions across all services within the network slice}, \\
Consider $U$ user services within a network slice. Let $u_1, u_2,...,u_{n-1}$ denote each user service. We define the total probability across all users services is,
\begin{equation}
\begin{split}
P(u_{1}) & + \sum_{n=2}^{U} P(u_{n}|u_{1},...,u_{n-1}) = \underbrace{\frac{1}{\sigma_{s}^{2}} e^{-\frac{x}{\sigma_{s}^{2}}}}_\text{$1^{st} \; term$} + \\ & \underbrace{\sum_{n=2}^{U}P(u_{n}|u_{1},...,u_{n-1})}_\text{$2^{nd} term$}
\end{split}
\end{equation}
\hspace{0.1cm}
where $\frac{1}{\sigma_{s}^{2}} e^{-\frac{x}{\sigma_{s}^{2}}}$ is the SNR distribution of first user in the network that follows Rayleigh fading (assumption). Since we are dealing with discrete and independent slice-in-slice category entities, we model the probability distributions across users services as Poisson distributions,
\begin{equation}
P(u_{1}, u_{2},..., u_{n}) = \frac{e^{-\lambda_{1}} (\lambda_{1})^{n}}{n!}
\end{equation}
\begin{equation}
P(u_{1}, u_{2},..., u_{n-1}) = \frac{e^{-\lambda_{2}} (\lambda_{2})^{n-1}}{(n-1)!}
\end{equation}
\hspace{0.1cm}
where $\lambda_{1}$ and $\lambda_{2}$ are expected valusers defined in general as,
where (generalized) $\lambda$ is the expectation which we define as,
\begin{equation}
\begin{split}
\lambda & = \sum_{i=1}^{n} i^{th} \; user \; service \; * P(i^{th} \; user \; service \; of \; \\ &  throughput \; u_{i}(t))
\end{split}
\end{equation}
\hspace{0.1cm}
Using (28), we define $\lambda_{1}$ and $\lambda_{2}$ as,
\begin{equation}
\begin{split}
\{\lambda_{1}, \lambda_{2}\} & = \{\frac{\sum_{i=1}^{n} u_{i}(t)}{c(t)}, \frac{\sum_{i=1}^{n-1} u_{i}(t)}{c(t)}\}
\end{split}
\end{equation}
\hspace{0.1cm}
where $u_{i}(t) = \frac{f_{d} e^{\beta r_{i} s_{i}}}{\Delta t}$ is the throughput [3] of each $i^{th}$ user service, $\beta$ is a utility convergance constant, $r_{i}$ is the system resource allocation, $s_{i}$ is the signal to noise ratio (SNR) and $c(t)$ is the cell throughput of network slice. 
 \begin{figure}
\begin{center}
\includegraphics[scale = 0.23]{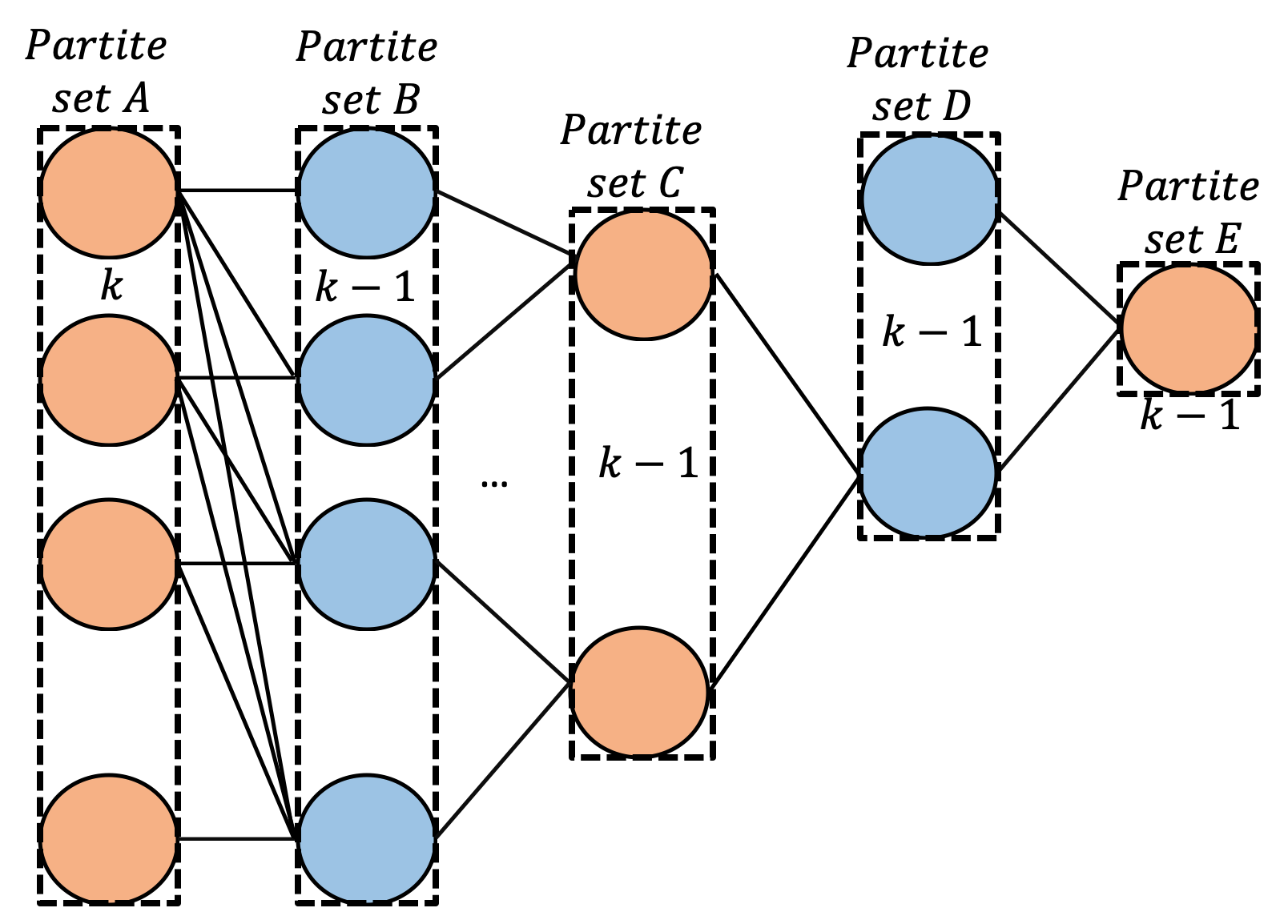}
\caption{Chromatic color polynomial for $k$-partite graphs}
\end{center}
\vspace{-5mm}
\end{figure}
Using (26), (27) and $2^{nd} \; term$ of (25), the conditional probability is,
\begin{equation}
\begin{split}
\frac{P(u_{1}, u_{2},...,u_{n-1},u_{n})}{P(u_{1}, u_{2},...,u_{n-1})} & =  e^{-\frac{f_{d}}{\Delta t}\frac{\sum_{i=1}^{n} i e^{\beta r_{i} s_{i}}}{c(t)}} \\ & (\frac{f_{d}}{\Delta t}\frac{\sum_{i=1}^{n} i e^{\beta r_{i} s_{i}}}{c(t)})^{n} \\ & \frac{1}{e^{-\frac{f_{d}}{\Delta t}\frac{\sum_{i=1}^{n-1} i e^{\beta r_{i} s_{i}}}{c(t)}} (\frac{f_{d}}{\Delta t}\frac{\sum_{i=1}^{n-1} i e^{\beta r_{i} s_{i}}}{c(t)})^{n-1} n}
\end{split}
\end{equation}
\hspace{0.1cm}
Inserting (30) in (25) we express the total probability distribution function,
\begin{equation}
\begin{split}
\frac{1}{\sigma_{s}^{2}} e^{-\frac{x}{\sigma_{s}^{2}}} +  \sum_{n=2}^{U} \frac{e^{-\frac{f_{d}}{\Delta t}\frac{\sum_{i=1}^{n} i e^{\beta r_{i} s_{i}}}{c(t)}} (\frac{f_{d}}{\Delta t}\frac{\sum_{i=1}^{n} i e^{\beta r_{i} s_{i}}}{c(t)})^{n}}{e^{-\frac{f_{d}}{\Delta t}\frac{\sum_{i=1}^{n-1} i e^{\beta r_{i} s_{i}}}{c(t)}} (\frac{f_{d}}{\Delta t}\frac{\sum_{i=1}^{n-1} i e^{\beta r_{i} s_{i}}}{c(t)})^{n-1} n}
\end{split}
\end{equation}
\hspace{0.1cm}
Thus, to estimate the system resources $r_{i}$, we define that the objective function needs to be maximized and differentiated with respect to the known entity, i.e., SNR ($s_{i}$),
\begin{equation}
max_{r_{i}} \; \frac{\partial(Eqn. (31))}{\partial s_{i}} = 0
\end{equation}
\hspace{0.1cm}

\section{Result 5}
\textbf{Result 5}: \textit{For a $n$-partite graph, the permutation coloring strategy uses $k(k-1)^{n-1}$ colors, where $k$ is the maximum available colors}. \\
A partite graph is defined as a graph where the vertcies can be partitioned and represented into $k$ disjoint and independent sets. Partite graphs are useful which can be used to represent and model several relationships, such as the probability distribution functions discussed in Result 4. Consider a $n-$partite graph shown in Fig. 2 defined as,
\begin{equation}
\independent v_{i} \in A, \independent v_{j} \in B, ..., \independent v_{k} \in E \; \forall \; i, j, k
\end{equation}
\hspace{0.1cm}
where $v_{i}, v_{j}, ..., v_{k}$ are the vertices $\in$ $A, B, ..., E$. We now define the total number of colors needed for coloring a $n $-partite graph. We know from the concept of graph coloring that no two adjacent vertices can be assigned the same color. It is observed that to color a $n-$ partite graph, only $2$ color combinations would be used (orange and blue) out of a total of $k$ available colors. This is because there is no direct relation (edges or lines) between vertices belonging to alternate partite sets, i.e., $\{A, C, E\}$ or $\{B, D\}$. Generalizing, we can express that $k$ colors would be used for coloring the first partite set (eg. $A$) and $k-1$ colors for coloring the subsequent $n-1$ partite sets (eg. $B, C, D, E$) (as the colors can be re-used for alternating ones). Thus, we define the chromatic polynomial for a $n-$partite graph as,
\begin{equation}
\mathcal{C} = \prod_{p=1} k(k-p)^{n-1}
\end{equation}

\section{Conclusion}
In this paper, we discussed and explained some of the useful novel optimization results to achieve 5G Slice-in-Slice Scheduling. ORAN Slicing combined with the optimization results can facilitate and introduce a new angle to implement challenging RAN SLA mechanisms. The results and approaches will thereby facilitate closed-loop optimization to generate and deploy models to control the real-time RAN operation.


\begin{thebibliography}{00}
\bibitem{b1} J. Tsitsiklis, Inference and Limit Theorems, The Chebyshev Inequality,  MITOpenCourseWare.
\bibitem{b1} D.B.West, Introduction to Graph Theory, Pearson Education (Singapore), Second edition, ISBN 81-78080830-3, 2002.
\bibitem{b1}  S. Chaudhuri, I. Baig, Debabrata Das, QoS aware Downlink Scheduler
for a Carrier Aggregation LTE-Advance Network with Efficient Carrier
Power Control, in IEEE INFOCOM, Bangalore, India, 2018.
\end{thebibliography}
\end{document}